\documentclass[twocolumn,showpacs,amsmath,amssymb,pra]{revtex4}

\usepackage{graphicx}
\usepackage{times}

\newcommand{\op}[1]{\hat{#1}}

\newcommand{\cre}[1]{\hat{#1}^\dagger}
\newcommand{\des}[1]{\hat{#1}}

\begin{document}

\title{Cooling and squeezing via quadratic optomechanical coupling}

\author{A.~Nunnenkamp}
\affiliation{Departments of Physics and Applied Physics, Yale University, New Haven, Connecticut 06520, USA}
\author{K.~B\o rkje}
\affiliation{Departments of Physics and Applied Physics, Yale University, New Haven, Connecticut 06520, USA}
\author{J.~G.~E.~Harris}
\affiliation{Departments of Physics and Applied Physics, Yale University, New Haven, Connecticut 06520, USA}
\author{S.~M.~Girvin}
\affiliation{Departments of Physics and Applied Physics, Yale University, New Haven, Connecticut 06520, USA}

\date{\today}

\begin{abstract}
We explore the physics of optomechanical systems in which an optical cavity mode is coupled parametrically to the \emph{square} of the position of a mechanical oscillator. We derive an effective master equation describing two-phonon cooling of the mechanical oscillator. We show that for high temperatures and weak coupling, the steady-state phonon number distribution is non-thermal (Gaussian) and that even for strong cooling the mean phonon number remains finite. Moreover, we demonstrate how to achieve mechanical squeezing by driving the cavity with two beams. Finally, we calculate the optical output and squeezing spectra. Implications for optomechanics experiments with the membrane-in-the-middle geometry or ultracold atoms in optical resonators are discussed.
\end{abstract}


\maketitle

\emph{Introduction.}
In optomechanical systems optical and mechanical degrees of freedom are coupled via radiation pressure, optical gradient, or photothermal forces. While work in this area was originally motivated by the goal of building sensitive detectors for gravitational waves \cite{Caves1980, Barish1999}, the field has become an active area of research in its own right. Its main goal is to investigate quantum coherence in macroscopic solid-state devices both for quantum information purposes and gaining new insights into the quantum-to-classical transition \cite{Kippenberg2008, Marquardt2009}.

In most optomechanical experiments an optical cavity mode is parametrically coupled to the position of a mechanical oscillator. Consequently, many properties of this setup have been discussed, including red-sideband laser cooling in the resolved-sideband limit \cite{Wilson-Rae2007, Marquardt2007}, normal-mode splitting \cite{Dobrindt2008, Groblacher2009}, optical squeezing \cite{Mancini1994, Fabre1994}, backaction-evading measurements \cite{Braginsky1980, Clerk2008, Hertzberg2010}, mechanical squeezing using either feedback \cite{Clerk2008, Woolley2008}, squeezed light \cite{Jahne2009} or modulation of input power \cite{Mari2009}, and entanglement between light and a mechanical oscillator \cite{Vitali2007}.

However, some optomechanical systems feature a quadratic optomechanical interaction, i.e.~an optical cavity mode is coupled parametrically to the \emph{square} of the position of a mechanical oscillator. One example is the membrane-in-the-middle geometry, in which the membrane is placed at a node or antinode of the cavity field \cite{Thompson2008, Jayich2008, Sankey2010}. A second system which can realize quadratic optomechanical coupling is a cloud of ultracold atoms loaded into an optical cavity, where the cloud's center-of-mass coordinate serves as the mechanical degree of freedom \cite{Murch2008, Purdy2010}. To date, the theoretical literature has focused on using the quadratic coupling to detect phonon Fock states \cite{Thompson2008, Jayich2008, Helmer2009, Miao2009, Clerk2010}, but otherwise the possible uses of this form of optomechanical coupling are largely unstudied.

In this paper we explore three features of quadratic optomechanical coupling: two-phonon cooling of the mechanical oscillator, squeezing of the mechanical oscillator, and squeezing of the optical output field. Using Fermi's Golden rule we first write down an effective master equation for the mechanical oscillator. In the classical limit of large phonon number, two-phonon cooling processes change the steady-state number distribution from exponential to Gaussian, a consequence of the nonlinear damping. In the quantum limit we find that ground-state cooling is not possible since two-phonon cooling processes preserve the phonon-number parity. We then demonstrate that the model maps onto a degenerate parametric oscillator if the cavity is driven by two laser beams whose frequencies are detuned to either side of the cavity resonance by an amount equal to the mechanical frequency. This opens up the possibility of mechanical squeezing. Finally, we calculate the optical output spectrum and find that this system is capable of producing optical squeezing.

\emph{Hamiltonian.}
We start from the Hamiltonian (with $\hbar = 1$)
\begin{equation}
\op{H} = \left(\omega_R + g \hat{x}^2 \right) \left(\cre{a}\des{a} - \langle \cre{a} \des{a} \rangle \right) + \omega_M \cre{b}\des{b} + \op{H}_\gamma + \op{H}_\kappa
\label{Hamiltonian}
\end{equation}
where $\omega_R$ is the cavity resonance frequency, $g$ the quadratic optomechanical coupling, and $\hat{x} = x_\mathrm{ZPF} (\des{b} + \cre{b})$ the position of the mechanical oscillator with zero-point fluctuations $x_\mathrm{ZPF} = (2m\omega_M)^{-1/2}$, frequency $\omega_M$ and mass $m$. $\des{a}$ and $\des{b}$ are annihilation operators obeying bosonic commutation relations. $\op{H}_\gamma$ and $\op{H}_\kappa$ describe the coupling to the mechanical and optical baths and the optical drive. We have subtracted the steady-state mean photon number $\langle \cre{a} \des{a} \rangle$ which renormalizes the frequency of the mechanical oscillator. The Hamiltonian (\ref{Hamiltonian}) is relevant to systems with membrane-in-the-middle geometry \cite{Thompson2008}, to ultracold atoms in optical resonators \cite{Murch2008, Purdy2010}, and double-microdisk whispering-gallery mode resonators \footnote{O.~Painter, private communication.} when the first derivative of the cavity dispersion relation $\omega_\mathrm{cav}(x)$ vanishes, i.e.~$\omega_\mathrm{cav}'(x_0)=0$, so that $g = \omega_\mathrm{cav}''(x_0)/2$ is the leading order of the optomechanical coupling.

Expressing $\hat{a} = e^{-i \omega_L t} (\bar{a} + \hat{d})$ with the laser frequency $\omega_L$, choosing $\bar{a}$ real, and neglecting $\cre{d}\des{d}$ with respect to $\bar{a}(\cre{d} + \des{d})$, we obtain the following quantum master equation
\begin{equation}
\dot{\varrho} = - i \left[ \op{H}_S, \varrho \right] + \kappa \mathcal{D}[\des{d}]\varrho + \gamma (1+n_\mathrm{th}) \mathcal{D}[\des{b}]\varrho + \gamma n_\mathrm{th} \mathcal{D}[\cre{b}]\varrho
\label{fullmastereqn}
\end{equation}
with the system Hamiltonian
\begin{equation}
\op{H}_S = - \Delta \cre{d}\des{d} + \omega_M \cre{b}\des{b} + \bar{g} (\des{b} + \cre{b})^2 (\cre{d} + \des{d})
\label{fullHam}
\end{equation}
where $\Delta = \omega_L - \omega_R$ is the detuning, $\bar{g} = g \bar{a} x_\mathrm{ZPF}^2$ the coupling, $\kappa$ and $\gamma$ the cavity and mechanical damping rates, and $n_\mathrm{th}$ the thermal phonon number. We assume that the optical bath is at zero temperature. $\mathcal{D}[\des{o}] \varrho = \des{o} \varrho \cre{o} - (\cre{o} \des{o} \varrho + \varrho \cre{o} \des{o})/2$ denotes the standard dissipator in Lindblad form.

Current experiments with the membrane-in-the-middle geometry work in the resolved-sideband limit at large thermal phonon number, i.e.~$\kappa = 10^5 \mathrm{Hz}$, $\omega_M = 10^6 \mathrm{Hz}$, $\gamma = 0.1 \mathrm{Hz}$, and $n_\mathrm{th} = 10^7$ at $T = 300 \mathrm{K}$, while the optomechanical coupling is small compared to the cavity linewidth $\bar{g}/\kappa = 10^{-5}$ \cite{Thompson2008, Jayich2008, Sankey2010}. Experiments with ultracold atoms in optical resonators have recently also realized quadratic optomechanical coupling, albeit with much larger coupling $\bar{g}/\kappa \approx 1$ at small thermal phonon number $n_\mathrm{th}$. However, they operate in the single-photon regime and outside the resolved-sideband limit, so that our results do not directly apply \cite{Murch2008, Purdy2010}.

\begin{figure}
\centering
\includegraphics[width=42mm]{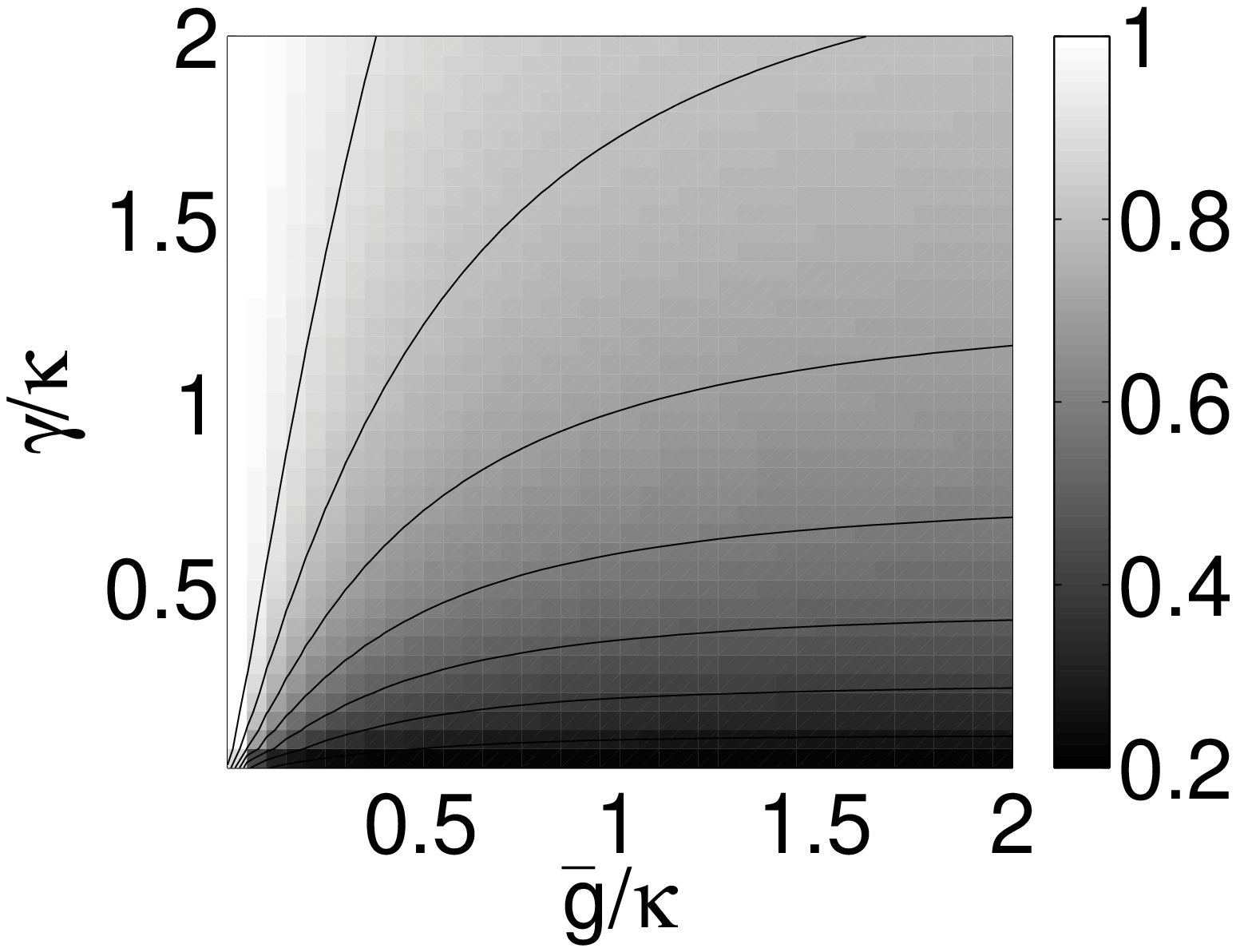}
\includegraphics[width=42mm]{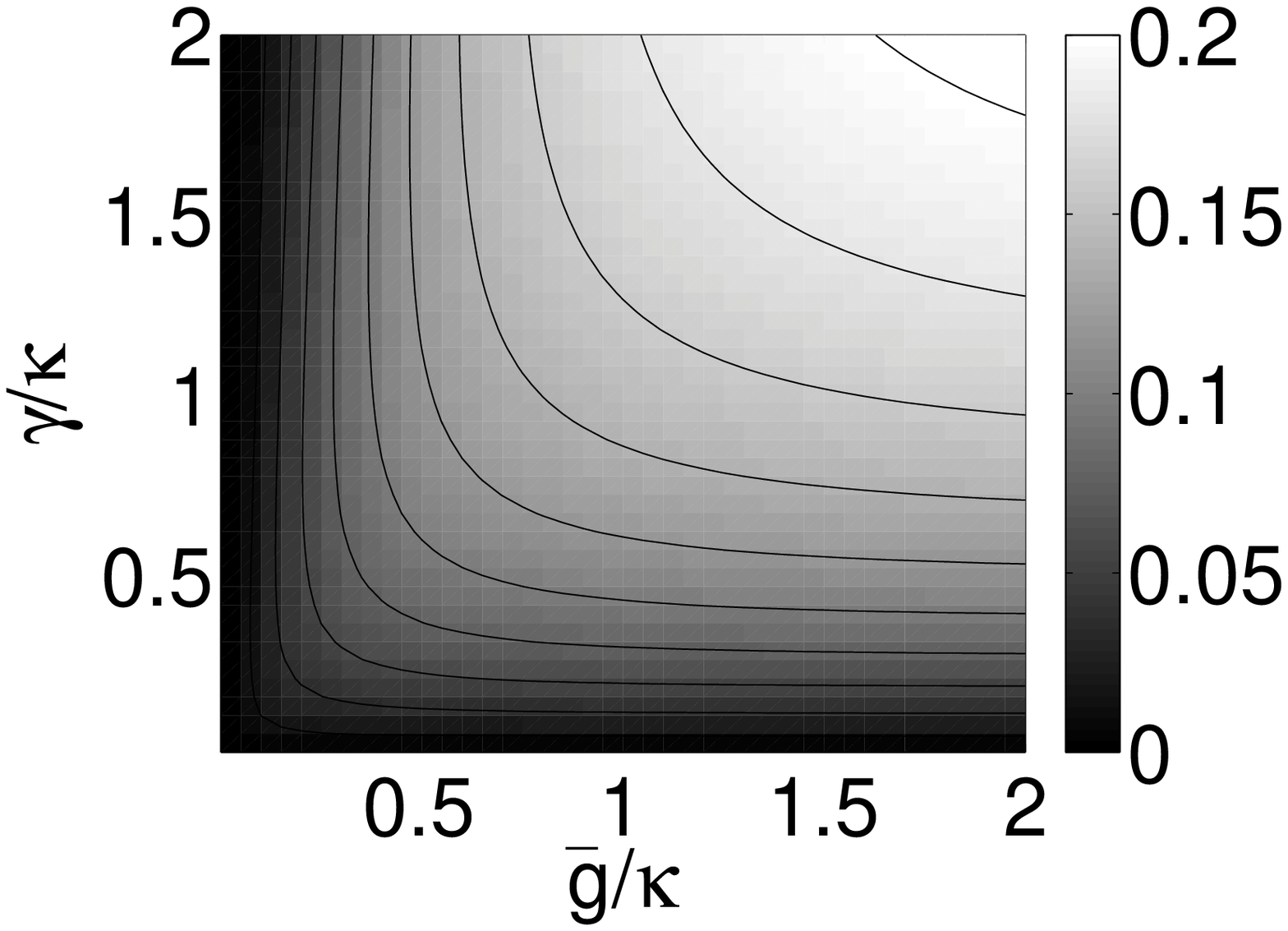}
\caption{Steady-state mean phonon $\langle \cre{b} \des{b} \rangle$ (left) and photon number $\langle \cre{d} \des{d} \rangle$ (right) as a function of the coupling strength $\bar{g}/\kappa$ and thermal coupling $\gamma/\kappa$ for thermal phonon number $n_\mathrm{th} = 1$, obtained from the numerical solution of the full quantum master equation (\ref{fullmastereqn}).}
\label{fig:phasediagram}
\end{figure}

\emph{Numerical simulations of the full quantum master equation.}
If the cavity is driven on the red two-phonon resonance, i.e.~$\Delta = -2\omega_M$, in the good-cavity limit, i.e.~$\kappa \ll \omega_M$, we expect two-phonon cooling processes to be important. Concentrating on the resonant terms, i.e.~$\hat{H}_S = \bar{g} (\cre{b}\cre{b}\des{d} + h.c.)$, we solve the full quantum master equation (\ref{fullmastereqn}) numerically for small thermal phonon numbers $n_\mathrm{th}$. As the coupling strength $\bar{g}$ increases, we find that (i) for $\gamma n_\mathrm{th} \ll \kappa$ the mechanical oscillator is cooled due to the coupling to the zero-temperature bath of the optical field \cite{GirvinRMP}, (ii) for $\gamma n_\mathrm{th} \gg \kappa$ the optical field is heated by the coupling to the mechanical oscillator, and (iii) for $\gamma n_\mathrm{th} \approx \kappa$ both effects are important and the density matrix has non-zero off-diagonal elements so that e.g.~the correlator $|\langle \cre{d} \des{b} \des{b} \rangle|$ is non-zero. These features are generic, but the details depend on the thermal phonon number $n_\mathrm{th}$. As an example we plot in Fig.~\ref{fig:phasediagram} the steady-state mean phonon number $\langle \cre{b} \des{b} \rangle$ and the mean number of photons due to the coupling to the membrane $\langle \cre{d} \des{d} \rangle$ as a function of the coupling strength $\bar{g}/\kappa$ and thermal coupling $\gamma/\kappa$ for $n_\mathrm{th} = 1$.

\emph{Effective master equation describing two-phonon cooling.}
For $\gamma n_\mathrm{th} \ll \kappa$ and weak coupling, i.e.~$\bar{g} \ll \kappa$, we can employ a quantum noise approach \cite{GirvinRMP}, i.e.~we calculate two-phonon cooling and amplification rates using Fermi's Golden rule. Concentrating on the diagonal terms of the density matrix $\varrho_{nn} = P_n$, we write down a set of rate equations
\begin{align}
\dot{P}_n &= - \gamma \left( n_\mathrm{th} (n+1) + ( n_\mathrm{th}+1) n \right) P_n \nonumber \\&
+ \gamma n_\mathrm{th} n P_{n-1} + \gamma ( n_\mathrm{th}+1) (n+1) P_{n+1} \nonumber \\ &
- \left( \Gamma_\downarrow n (n-1) + \Gamma_\uparrow (n+2) (n+1) \right) P_n \nonumber \\ &
+ \Gamma_\downarrow (n+2) (n+1) P_{n+2} + \Gamma_\uparrow n (n-1) P_{n-2}.
\label{rateeqn}
\end{align}
The terms in the first two lines are due the coupling of the mechanical oscillator to its thermal bath with rate $\gamma$ and thermal phonon number $n_\mathrm{th}$. The terms in the last two lines are due to two-phonon processes whose rates are governed by $\Gamma_{\downarrow/\uparrow} = g^2 x_\mathrm{ZPF}^4 S_{nn}(\pm2\omega_M)$ where $S_{nn}(\omega) = \kappa |\bar{a}|^2 |\chi_R(\omega)|^2$ is the photon number spectral density and $\chi_R(\omega) = [\kappa/2 - i(\omega+\Delta)]^{-1}$ the cavity response function. We note that nonlinear damping of an oscillator has been studied in Ref.~\cite{Dykman1975, Dykman1978}.

The infinite set of rate equations (\ref{rateeqn}) can be replaced by a single differential equation for the generating function $F(z,t) = \sum_{n=0}^\infty P_n(t) z^n$. For vanishing two-phonon amplification $\Gamma_\uparrow = 0$ the steady-state equation can be solved exactly in terms of the confluent hypergeometric function \cite{Dodonov1997}.

We point out that it is an advantage of quadratic cooling that it enables cooling when the membrane or atoms are placed at a node of the cavity field. In this situation the system is most insensitive to absorption from the membrane or atoms as well as to bistability from radiation pressure.

\emph{Two-phonon cooling in the classical limit.} For large thermal phonon number $n_\mathrm{th} \gg 1$ and weak two-phonon cooling $\gamma n_\mathrm{th} \gg \Gamma_\downarrow$ we can replace the quantum operators $\hat{b}$ and $\hat{d}$ in their Heisenberg equations of motion by complex amplitudes $\beta = \langle \hat{b} \rangle$ and $\alpha = \langle \hat{d} \rangle$ and obtain two coupled classical Langevin equations $\dot\beta = - \gamma \beta/2  - 2 i \bar{g} \beta^* \alpha + \xi$ and $\dot\alpha = - \kappa \alpha/2  - i \bar{g} \beta^2$, where the thermal noise is characterized by $\langle \xi(t) \xi^*(t') \rangle = \gamma n_\mathrm{th} \delta(t-t')$. Adiabatically eliminating the optical field we obtain an equation of motion with a nonlinear damping term $\dot\beta = - \gamma \beta/2 - 4 \bar{g}^2 |\beta|^2 \beta/\kappa + \xi$. Solving the corresponding Fokker-Planck equation we obtain the phonon number distribution
\begin{equation}
P_n \propto \exp \left( -\frac{n}{n_\mathrm{th}} \right) \exp \left( -\frac{\Gamma_\downarrow n^2}{\gamma n_\mathrm{th}}\right).
\label{hightemp}
\end{equation}
The distribution changes from an exponential for $\gamma/\Gamma_\downarrow \gg n_\mathrm{th}$ to a Gaussian for $\gamma/\Gamma_\downarrow \ll n_\mathrm{th}$. In the latter limit the mean phonon number is given by $\langle n \rangle  = \sqrt{\gamma n_\mathrm{th}\kappa/\pi \bar{g}^2}$. The expression in Eq.~(\ref{hightemp}) agrees with the high-temperature limit of the exact solution to the rate equations \cite{Dodonov1997}. We conclude that the change in the steady-state phonon number distribution is a purely classical effect due to nonlinear damping.

\begin{figure}
\centering
\includegraphics[width=42mm]{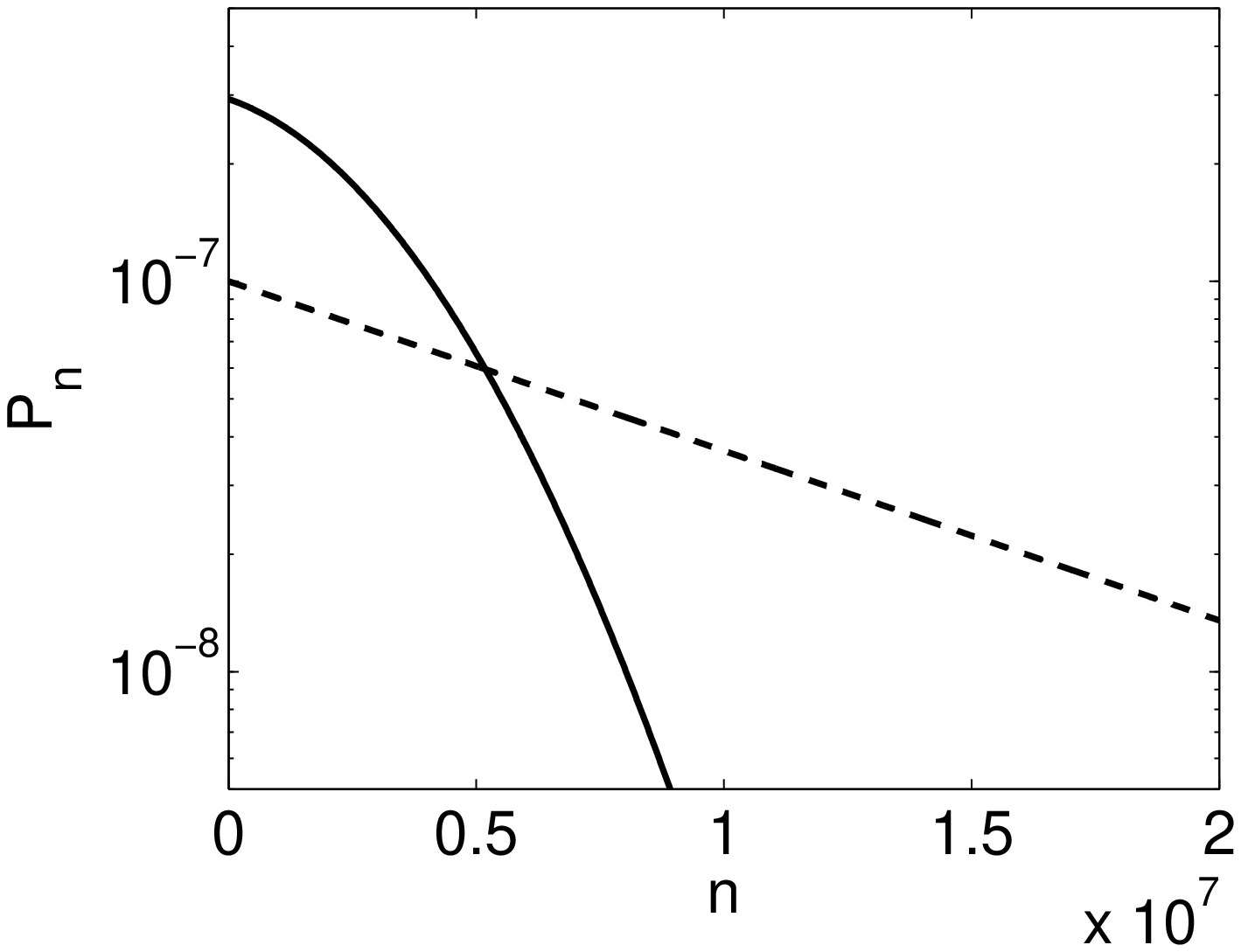}
\includegraphics[width=42mm]{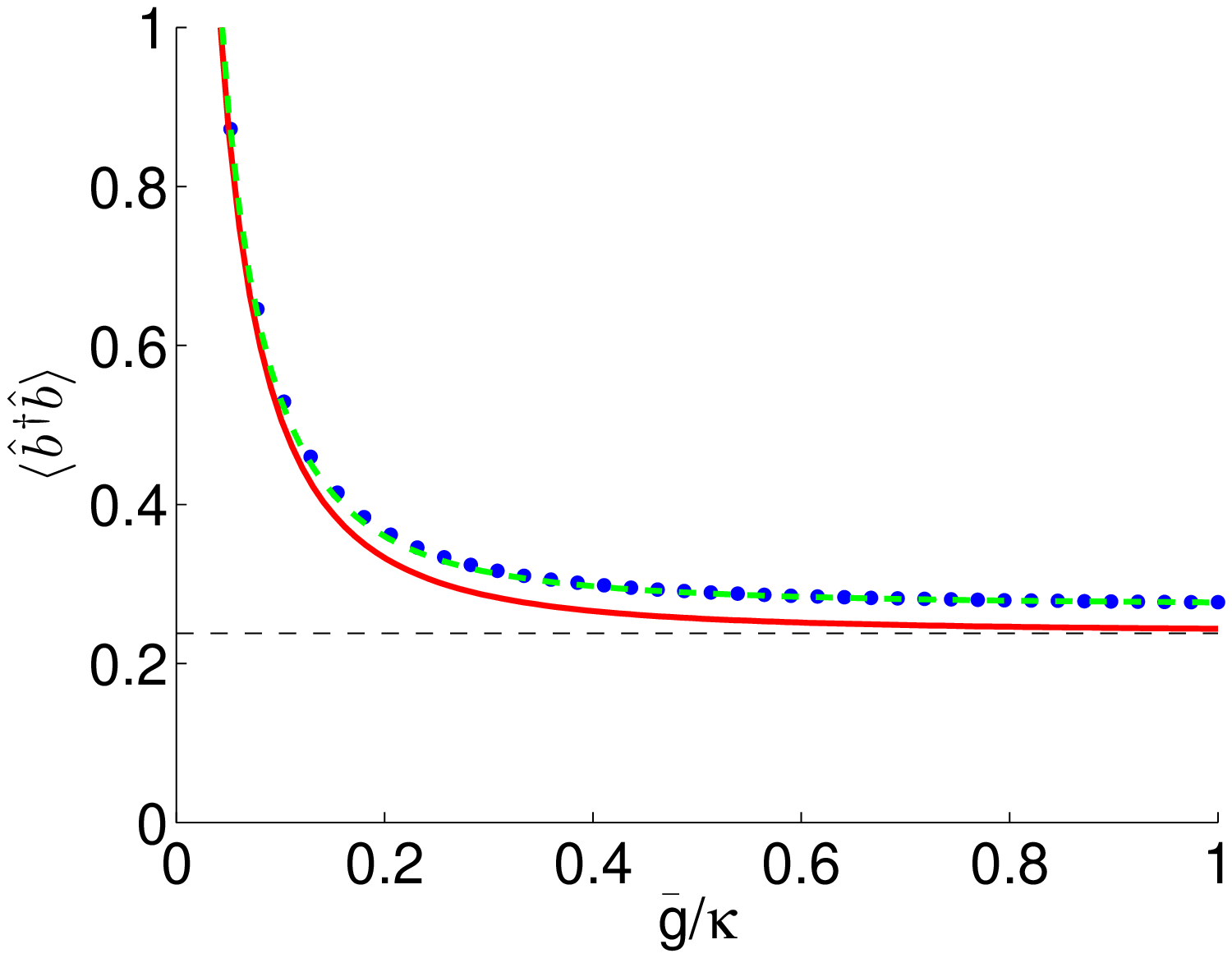}
\caption{
(Left) \emph{Classical limit:} Semi-logarithmic plot of the steady-state phonon number distribution $P_n$ for thermal phonon number $n_\mathrm{th} = 10^7$. Planck distribution for $\Gamma_\downarrow = 0$ (dashed) and nearly-Gaussian distribution for $\Gamma_\downarrow/\gamma = 4\cdot10^{-7}$ (solid).
(Right) \emph{Quantum limit:} Mean phonon number $\langle \cre{b} \des{b} \rangle$ as a function of coupling $\bar{g}/\kappa$ obtained from the quantum master equation (blue dots) and the classical rate equations without (red solid) and with strong-coupling correction (green dashed). The black dashed line indicates the minimal phonon number $\bar{n}_M^0$ for $n_\mathrm{th} = 5$.}
\label{fig:coolinghigh}
\label{fig:coolinglow}
\end{figure}

The quadratic coupling in current membrane-in-the-middle experiments is very small. Nonetheless, it leads to sizable effects if the thermal phonon number $n_\mathrm{th}$ is large. In Fig.~\ref{fig:coolinghigh} (left) we plot the steady-state phonon number distribution for a thermal phonon number $n_\mathrm{th} = 10^7$, corresponding to a temperature of $T = 300 \mathrm{K}$, both in the absence $\Gamma_\downarrow = 0$ and presence $\Gamma_\downarrow/\gamma = 4\cdot10^{-7}$ of two-phonon cooling. We find the Planck distribution with mean $n_\mathrm{th} = 10^7$ becomes a nearly-Gaussian distribution with mean $n_\mathrm{th} = 2.8 \cdot 10^6$. The qualitative change in the phonon number distribution for membrane-in-the-middle experiments is the main result of this paper.

\emph{Two-phonon cooling in the quantum limit.}
Solving the rate equations (\ref{rateeqn}) in the limit of strong optical damping and/or  small thermal heating rate $\gamma n_\mathrm{th}$, i.e.~$\Gamma_\downarrow \gg \gamma n_\mathrm{th}$, we obtain the phonon number distribution $P_0/P_1 = 3 + 1/n_\mathrm{th}$ with all other $P_n = 0$ and the minimal mean phonon number $\bar{n}_M^0 = 1/(4 + 1/n_\mathrm{th})$.
The fact that strong cooling leaves both ground and first excited state occupied is a consequence of the fact that two-phonon cooling processes preserve the phonon-number parity.

In Fig.~\ref{fig:coolinglow} (right) we plot the steady-state phonon number $\langle \cre{b} \des{b} \rangle$ as a function of coupling $\bar{g}$ obtained from the full quantum master equation (\ref{fullmastereqn}) and the rate equations (\ref{rateeqn}). The two results show excellent agreement at weak coupling $\bar{g} \ll \kappa$.
As $4\bar{g}^2/\kappa$ becomes comparable to $\kappa$, the Fermi's Golden rule expression for two-phonon cooling breaks down and the predictions of the full master equation (\ref{fullmastereqn}) and the weak-coupling rate equations (\ref{rateeqn}) will in general be different. In the limit where only the Fock states with phonon number $n \le 3$ are important, we adiabatically eliminate the off-diagonal terms in the quantum master equation (\ref{fullmastereqn}) and find that the strong-coupling two-phonon cooling rates are given by
\begin{equation}
\Gamma^{\downarrow}_{n,n-2} = \frac{4\bar{g}^2 n (n-1) \kappa}{\kappa^2 + 4\bar{g}^2 n (n-1)}.
\end{equation}
In the weak-coupling limit $\bar{g} \ll \kappa$ this expression simplifies to our previous result $4\bar{g}^2 n (n-1) /\kappa$. In the limit of strong coupling $\bar{g} \gg \kappa$, the two-phonon cooling rate $\Gamma^{\downarrow}_{n,n-2}$ remains finite and cannot exceed the cavity damping rate $\kappa$. This leads to a minimum phonon number which is larger than the one predicted from the weak-coupling theory. In Fig.~\ref{fig:coolinglow} (right) we plot the steady-state mean phonon number obtained from the modified rate equations and find excellent agreement with the exact solution to the full quantum master equation (\ref{fullmastereqn}).

We note that for quadratic coupling in the good-cavity limit, the phonon number distribution $P_n$ can be measured by monitoring the phase shift of the reflected light \cite{Thompson2008, Jayich2008, Helmer2009}.

\emph{Mechanical amplification and squeezing.}
Driving the cavity at both $\omega_R \pm \omega_M$ with equal strength, the classical part of the cavity field oscillates in time $\bar{a} = A \cos \omega_M t$. In the case of linear optomechanical coupling this enables a backaction-evading measurement of one quadrature of the mechanical oscillator \cite{Braginsky1980, Clerk2008, Hertzberg2010}, but does not itself produce squeezing apart from the one which is conditioned on the measurement outcome.
When we instead consider the same drive applied to a system with quadratic coupling, moving to an interaction picture with respect to $\op{H}_0 = \omega_R \cre{a}\des{a} + \omega_M \cre{b}\des{b}$ and keeping only non-rotating terms of the quadratic coupling, we obtain the standard Hamiltonian of the degenerate parametric oscillator, i.e.~$\op{H}_\textrm{DPO} = \frac{\chi}{2} ( \des{b}^2 + (\cre{b})^2 )$ with $\chi = g x_\mathrm{ZPF}^2 A/2$. Thus this setup could be used to amplify small mechanical signals \cite{GirvinRMP}.

Solving the linear quantum Langevin equations we see that the steady-state fluctuations in the quadrature $\hat{X}=(\hat{b}e^{i \omega_M t}+ H.c.)/\sqrt{2}$ of the mechanical oscillator are squeezed below the thermal level, $\langle \hat{X}^2 \rangle = \frac{n_\mathrm{th} + 1/2}{1 + 2 \chi / \gamma}$, depending on the ratio $\chi/\gamma$. At threshold $\chi = \gamma/2$, i.e.~before the parametric oscillator becomes unstable, the squeezing reaches the theoretical limit for an internal mode of 3dB \cite{Milburn1981}. Although the parameters of current membrane-in-the-middle setups show small coupling $\bar{g} \ll \kappa$ and large thermal phonon number $n_\mathrm{th} \gg 1$, we emphasize that the ratio $\chi/\gamma$ which is of importance here can still be comparable to unity and lead to significant noise squashing.

The mechanical squeezing can be detected by coupling the position of the mechanical oscillator parametrically to a second optical mode. This scenario has been studied in Ref.~\cite{Woolley2008}.

\begin{figure}
\centering
\includegraphics[width=86mm]{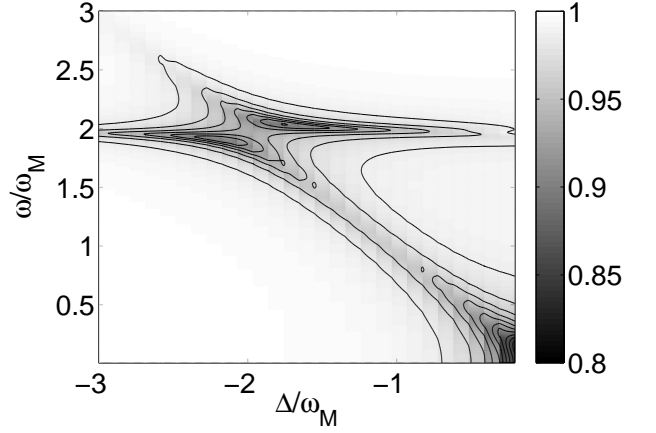}
\caption{Optimal squeezing spectrum of the cavity output field $S^\mathrm{out}_\mathrm{opt}(\omega)$ as a function of detuning $\Delta$. The parameters are $\omega_m/\kappa = 5$, $\bar{g}/\kappa = 0.5$, $\gamma/\kappa = 0.1$ and $n_\mathrm{th} = 0$.}
\label{fig:squeezing}
\end{figure}

\emph{Output spectrum.}
Let us now return to the full model (\ref{fullHam}). Up to second order in the coupling $\bar{g}$, the cavity output spectrum which is defined as $S^\mathrm{out}_{dd}(\omega) = \int dt \, e^{i \omega t} \langle \cre{d}_\mathrm{out}(t) \des{d}_\mathrm{out}(0) \rangle $ with $\des{d}_\mathrm{out} = \des{d}_\mathrm{in} + \sqrt{\kappa} \des{d}$ and $\langle \des{d}_\mathrm{in}(t) \cre{d}_\mathrm{in}(t) \rangle = \delta(t-t')$ is given by $S^\mathrm{out}_{dd}(\omega)= 4 \kappa \bar{g}^2 |\chi_R[-\omega]|^2 S_{x^2 x^2}(\omega)$ where
\begin{align}
S_{x^2x^2}(\omega) & = \frac{\gamma (n_\mathrm{th}+1)^2} {\gamma^2 + (\omega - 2 \omega_M)^2} \nonumber \\ & 
+ \frac{\gamma n_\mathrm{th}^2} {\gamma^2 + (\omega + 2 \omega_M)^2} + \frac{2\gamma n_\mathrm{th}(n_\mathrm{th}+1)}{\gamma^2 + \omega^2}.
\end{align}
We see the output spectrum $S^\mathrm{out}_{dd}(\omega)$ has sidebands at $\omega = \pm 2 \omega_m$ and $\omega = 0$ as expected for quadratic coupling.

\emph{Optical squeezing spectrum.}
One application of optomechanical devices which has been widely advocated \cite{Mancini1994, Fabre1994} is to use the nonlinear coupling between light and mirror to squeeze the incoming coherent light beam, i.e.~reduce one of its quadratures below the shot-noise level at certain frequencies. The quantity which characterizes this noise reduction is the optical squeezing spectrum $S_\theta^\mathrm{out}(\omega)$ given by \cite{Walls2008}
\begin{align}
S_\theta^\mathrm{out}(\omega) &= 1 + \int_{-\infty}^\infty dt \, e^{i \omega t} \langle :\hat{X}_\theta^\mathrm{out}(t),\hat{X}_\theta^\mathrm{out}(0) : \rangle  \nonumber \\ &
= 1 + \kappa \int_{-\infty}^\infty dt \, e^{i \omega t} \mathcal{T} \left[ \langle :\hat{X}_\theta(t),\hat{X}_\theta(0) : \rangle \right]
\end{align}
where $\hat{X}_\theta^\mathrm{out} = (\cre{d}_\mathrm{out} e^{i \theta} + H.c.)/2$, $\langle A,B \rangle = \langle A B \rangle - \langle A \rangle \langle B \rangle$, the colons indicate normal ordering and $\mathcal T$ is the time-ordering operator \cite{Walls2008}. The former expression is useful for calculations in input-output theory and the latter for master equation simulations evoking the quantum regression theorem.

In Fig.~\ref{fig:squeezing} we plot the optimal squeezing spectrum of the cavity output field $S^\mathrm{out}_\mathrm{opt}(\omega) = \min_\theta S_\theta^\mathrm{out}(\omega)$ as a function of detuning $\Delta$. In the good-cavity limit at zero temperature we find squeezing at $\omega = 0$ and $\omega = 2 \omega_M$ for $\Delta = 0$ and $\Delta = -2 \omega_M$, respectively. We choose parameters $\omega_m/\kappa = 5$, $\bar{g}/\kappa = 0.5$, $\gamma/\kappa = 0.1$ and $n_\mathrm{th} = 0$, possibly relevant to future experiments with ultracold atoms in optical resonators.

To gain insight beyond numerics we obtain the squeezing spectrum perturbatively up to second order in the coupling $\bar{g}$
\begin{align}
S_\mathrm{opt}^\mathrm{out}(\omega) &= 1 + S^\mathrm{out}_{dd}(\omega) + S^\mathrm{out}_{dd}(-\omega) \nonumber \\ &
- 8 \bar{g}^2 \kappa \Big| \chi_R(\omega) \chi_R(-\omega) S_{x^2 x^2}(\omega) \nonumber \\ &
\left. - \left( n_\mathrm{th} + \frac{1}{2} \right) \chi^*_R(\omega) \chi_R(-\omega) \left( \kappa \chi_R(\omega)-1 \right) \right. \nonumber \\ & 
\times \left. \left[\frac{1}{\gamma+i(\omega-2\omega_M)} - \frac{1}{\gamma+i(\omega+2\omega_M)} \right] \right|.
\end{align}
As in the case of linear optomechanical coupling \cite{Mancini1994, Fabre1994}, we find two regimes of squeezing: for small detuning and small frequencies as well as for detuning and frequencies close to twice the mechanical frequency. For the same parameters and at zero temperature the amount of squeezing is comparable to the linear optomechanical systems. However, thermal fluctuations are more destructive for squeezing effects with quadratic optomechanical coupling due to the quadratic scaling with the thermal phonon number $n_\mathrm{th}$ and due to the fact that the spectrum $S_{x^2x^2}(\omega)$ has weight at zero frequency.

\emph{Conclusions.}
We have explored the physics of nonlinear optomechanical systems where an optical cavity mode couples quadratically rather than linearly to the position of a mechanical oscillator. For optomechanical experiments with membrane-in-the-middle geometry, we predict a qualitative change in the phonon number distribution and mechanical noise squashing. For future experiments with ultracold atoms in optical resonators, we found the quantum limit of two-phonon cooling as well as mechanical and optical squeezing.

\emph{Acknowledgements.} AN thanks E.~Ginossar for useful discussions. We acknowledge support  from NSF under Grant No.~DMR-0603369 (AN and SMG), DMR-0653377 (AN, JGEH, and SMG) and DMR-0855455 (JGEH), from the Research Council of Norway under Grant No.~191576/V30 (KB), and from AFOSR under Grant No.~FA9550-90-1-0484 (JGEH). This material is based upon work supported by DARPA under award No.~N6601-09-1-2100 and W911NF-09-1-0015. Part of the calculations were performed with the Quantum Optics Toolbox \cite{Tan1999}.

\end{document}